\documentclass[preprintnumbers,amsmath,amssymb,onecolumn]{revtex4}
\usepackage[dvips,colorlinks=true,citecolor=red,filecolor=green,linkcolor=blue,pdfnewwindow=true]{hyperref}
\usepackage{graphicx}
\usepackage{makeidx}
\usepackage{bm}
\usepackage[title]{appendix}

\def\epsilon{\varepsilon}

\def\beqr{\begin{eqnarray}}
\def\eqnr{\end{eqnarray}}
\def\beq{\begin{equation}}
\def\bc{\begin{center}}
\def\ec{\end{center}}
\def\eqn{\end{equation}}
\begin{document}
\title{Dynamics of unfolded protein aggregation}
\author{Utkarsh Upadhyay, Chandrima Barua, Shivani Devi, Jay Prakash Kumar and R.K. Brojen Singh}
\email{brojen@jnu.ac.in (Corresponding author)}
\affiliation{School of Computational \& Integrative Sciences, Jawaharlal Nehru University, New Delhi 110019, India.}
\date{\today}

\begin{abstract}
{\noindent}Unfolded protein aggregation in cellular system is a problem causing various types of diseases depending on which type unfolded proteins aggregate. This phenomenon of aggregation may take place during production, storage, shipment or delivery in the cellular medium. In the present work, we studied a simplified and extended version of unfolded protein aggregation model by Lumry and Eyring \textit{[J. Phys. Chem. 58:110–120 (1954).]} using stochastic approach. We solved analytically the Master equation of the model for the probability distribution $P(x,t)$ of the unfolded protein population and the solution was found to be time dependent complex binomial distribution. In the large population limit $P(x,t)\sim \Lambda(x,t)\times Pois(x,t)$. Further, the distribution became Normal distribution at large population and mean of the distribution limit: $P(x,t)\sim\Lambda(x,t)\times N(\langle qx\rangle,\langle qx\rangle)$. The fluctuations inherent in the dynamics measured by Fano factor can have sub-Poisson, Poisson and super-Poisson at different situations.\\

\end{abstract}

\maketitle

\noindent\textbf{\large Introduction}\\
Aggregation  of unfolded proteins could be a nuisance factor and lethal in many cellular dynamics causing various pathological states \cite{Fink,Albert}. This aggregation may lead to few important factors which might enhance potential immunogenicity \cite{Singh}, may cause conformational diseases (aggregation of mutant proteins) \cite{Gow} and neurodegenerative diseases (aggregation of prions in brain) \cite{Stefani}. The aggregation of unfolded proteins or non-native aggregation is a process of clustering together of monomers to stable complexes. Individual monomer could be composed of a single folded chain or multiple protein chains that are disulfide bonded to one another such as multimeric complex \cite{Roberts}. Few key features which favor protein aggregates are: the process is irreversible i.e. do not easily dissociate, and the process retains certain fraction of their original structure\cite{Joubert}. Active research in this area has grown up fast because of few reasons, first, unfolded protein aggregation is risk to develop immune system, second, in becoming potential key drug target for possible therapeutic intervention \cite{Jiskoot}, and these aggregated unfolded proteins can be possible key drug target for \cite{Martinez}. However, general theory of protein aggregation and control mechanism are still open question. \\

{\noindent}Unfolded protein aggregation may happen during production, storage, shipment or delivery to the patient. The process is subjected to various fluctuations (temperature, light, shaking, surfaces, pH adjustments, etc.), and these fluctuations favor protein aggregation in the cellular environment \cite{Chi}. There are many examples of protein aggregation, silicone oil droplets \cite{Tyagi} and freezing can induce aggregation \cite{Bhatnagar}. Further, this aggregation process is quite similar to the protein folding mechanism and the whenever, hydrophobic groups of a protein are exposed to the solvent, kinetic competition arises between folding and aggregation \cite{Mahler}. First step of the aggregation process could be to allow the the unfolded or partially folded protein in the process to choose a state either to move to native state or formation of dimer together with another unfolded molecule (aggregation) \cite{Zett}. This aggregation kinetics, structure and formation could be dependent on various factors which leads to the cause various diseases, for example, aggregation of amyloid-$\beta$ is believed to be cause of a number of brain diseases, including Alzheimer’s disease and Huntington’s disease \cite{Dobson}. Further,  the aggregation was found to be linked with induction of allergic responses, type 1 hypersensitivity responses, such as urticaria, anaphylaxis etc \cite{Mazzeo}. Sometimes such aggregation of proteins may lead to protein modifications from its own normal situation which cause neuronal dysfunction and neurotoxicity directed  to widespread neurodegeneration \cite{Tarawneh}.\\

{\noindent}Another aspect of unfolded protein aggregation is that it may occur throughout the lifetime of a protein causing modification in size, shape, morphology, chemical modifications, degradation and folding mechanism \cite{Moussa}. Such proteins in aggregated state generally have different biological functions as proteins at native states do causing toxic effects in the cellular systems \cite{Mahler}. Since slight  aggregation levels for certain period of time may cause clinically unacceptable situation,  such aggregation should be avoided \cite{Braun}. One possible strategy to prevent aggregation is to add molecules that hinders aggregation. For example, molecules that slow down aggregation urea, guanidinium chloride, amino acids, sugars, polyols, polymers, surfactants, and antibodies \cite{Cleland,Cleland1}. One reason causing this could be the decrease in the binding free energy of the unfolded protein and transition states allowing to increase in the aggregation activation energy slowing down aggregation rate. Sugars (sucrose and other molecules) that are preferentially excluded from the protein-solvent interface generally favors proteins to be in native state against aggregation \cite{Lee,Arakawa} leading to increase in free energy barrier of unfolding states \cite{Kendrick}. However, even though we know unfolded protein aggregation could be a potential target for prevention/cure of various types diseases, choice of molecules which can disrupt such aggregation state is a field of active research now.\\

{\noindent}We present a extended model of unfolded protein aggregation model by Lumry and Eyring \cite{Lumry} within stochastic formalism to understand dynamics and distribution of unfolded proteins. W solved exactly the master equation of the system for probability distribution of the unfolded proteins. Then we estimated noise fluctuation in the dynamics and analyzed the role of fluctuations in regulating the protein aggregation mechanism. We concluded few findings based on the results we obtained.\\

\noindent\textbf{\large Theory of protein aggregation model and solution}\\
Protein aggregation is problem in cellular process as the biological activity of aggregated proteins is not the same as that of native protein, and cause acute toxic in the cellular environment. Hence, one has to device strategies to prevent protein aggregation and is a subject of active research in pharmaceutic companies and biotechnological research. The number of proteins per human cell is roughly $\sim 10^{10}$ \cite{Milo} which are subjected to various forms of stress and a small fraction of marginally stable folded proteins in such stress and crowded environment cause error in folding leading to protein misfolding \cite{Dobson}. This process of misfolding is irreversible to folded state and cause the generation of unfolded protein aggregation \cite{Vendruscolo}. If $N$, $U$ and $A$ denote natured, denatured and aggregated proteins in the simple model of protein aggregation, the model can be described by the following set of reactions which is the extension of the model due to \cite{Lumry},
\begin{eqnarray}
\label{reaction}
N\stackrel{k_1}{\rightarrow}U;~U\stackrel{k_2}{\rightarrow}N;~U\stackrel{k_3}{\rightarrow}A;~A\stackrel{k_4}{\rightarrow}U
\end{eqnarray}
where, $k_1, k_2, k_3, k_4$ are rate constants. Using mass action chemical kinetics, this set of reactions \eqref{reaction} can be translated into the following set of ordinary differential equations,
\begin{eqnarray}
\frac{dN}{dt}&=&-k_1N+k_2U\nonumber\\
\frac{dU}{dt}&=&k_1N-k_2U-k_3U+k_4A\\
\frac{dA}{dt}&=&k_3U-k_4A\nonumber
\end{eqnarray}
Summing over these three differential equations we have, $\frac{dN}{dt}+\frac{dU}{dt}+\frac{dA}{dt}=0$ which implies $N+U+A=constant$. It has been reported that at room temperature the rate of folding protein is comparatively much higher than the rate of misfolding protein \cite{Mayor}. Hence, let us consider the dynamics of unfolded protein as few of large population of $N$ which move to unfolded protein aggregation state such that $U+A=constant-N\sim constant$. It has been reported that misfolded protein aggregation could be stable and long life span \cite{Roberts} and hence, one can take $A$ as constant or $A=constant-U$ as the mean length of the aggregated unfolded protein during exponential growth is nearly constant \cite{Kunes}.\\
\begin{figure}
\centering
\includegraphics[scale=0.4]{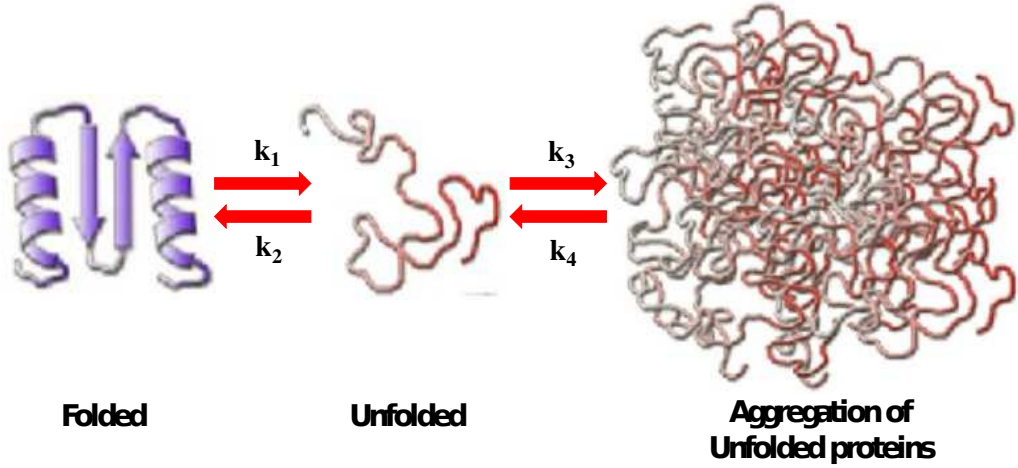}.
\caption{Schematic diagram illustrating the basic mechanism of the possible transitions of protein states in unfolded protein aggregation model. $\{k_i\}$, $i=1,2,3,4$ are the rates at which the transitions of states take place.}
\end{figure}

{\noindent}The role of protein aggregation can be studied if we obtain the distribution of unfolded protein $U$ from which protein aggregation takes place during the course of interaction. Let us take the population of $U$ at any instant of time '$t$' be $x$. The probability distribution of $x$ at time $t$ $P(x,t)$ can be calculated from the Master equation built from the reaction set \eqref{reaction}. The Master equation of the reactions set in \eqref{reaction} can be constructed as follows \cite{McQuarrie,Gillespie}. First, the stochastic rate constants are calculated using $c_i=V^{1-\nu_i}k_i$ \cite{Gillespie}, where, $\nu_i$ is the ith state change or stoichiometric parameter and since the reactions are all unimolecular reactions $\nu_i=1$ such that $c_i=k_i$. Then transition probabilities of each reaction are calculated by assuming the process is Markovian in forward arrow of time and taking all possible molecular interaction during time interval $[t,t+\Delta t]$ unto first order approximation, $\omega_i=f_i(x,N,A)\Delta t+O\left(\Delta^2\right)$, where, $f_i$ is the function obtained from the notion of molecular interaction \cite{McQuarrie,Gillespie}. The three variable model can be reduced to simple one dimension $f(x,t)$ by assuming $x\langle\langle N$ ($x$ is moving in the sea of $N$), and $x\rangle A$ (protein aggregation is fast decay). Then one can construct the Master equation by incorporating the above mentioned processes as given below,
\begin{eqnarray}
\label{master}
\frac{\partial P(x,t)}{\partial t} = (k_1N + k_4A)P(x-1,t) + (k_2 + k_3)(x+1)P(x+1,t) + (k_1N + k_2x + k_3x + k_4A)P(x,t)
\end{eqnarray}
This Master equation \eqref{master} can be solved using generating function technique defined by the transformation of $P(x,t)$ from $x$-space to $s$-space given by, $G(s,t)=\displaystyle\sum_{x}s^xP(x,t)$ provided initial condition $G(s,0)=s^M$ (initially there were $M$ population of $x$ at $t=0$) and normalization condition. Now multiplying equation \eqref{master} by $s^x$ and doing the summation over $x$ in both sides of the equation and after some algebra, we have the following differential equation in $G(s,t)$,
\begin{eqnarray}
\label{gf}
\frac{\partial G}{\partial t}+(k_2+k_3)(s-1)\frac{\partial G}{\partial s}= (s-1)(k_1N + k_4A)G
\end{eqnarray}
This partial differential equation in $G(s,t)$ eqref{gf} can be be used using Lagrange-Charpit characteristic method \cite{Delgado}. This can be done by writing the partial differential equation \eqref{gf} as in the following form,
\begin{eqnarray}
\label{lc}
\frac{\partial G}{(s-1)(k_1N + k_4A)G}  = \frac{\partial s}{(s-1)(k_3 + k_2)} = \frac{\partial t}{1}
\end{eqnarray}
By integrating both sides of the last equation in \eqref{lc} connecting $\partial s$ and $\partial t$, $(k_3 + k_2)\displaystyle\int\partial t= \int\frac{\partial s}{s-1}+C_1$, we arrived at the relation between $s$ and $t$ as, $\displaystyle\lambda_1 = (s-1)e^{-(k_2 + k_3)t}$, where, $\displaystyle\lambda_1=e^{C_1}$ is a constant in terms of integrating constant $C_1$. Next, from the equation connecting $\partial G$ and $\partial s$ in equation \eqref{lc}, by integrating both sides and rearranging the terms we get, $\displaystyle\int\frac{\partial G}{G}=\frac{k_1N + k_4A}{k_2 + k_3}\int\partial s+C_2$ and solution is $\displaystyle G=\lambda_2 e^{\frac{k_1N + k_4A}{k_2 + k_3}s}$, where, $\lambda_2$ is a constant in terms of integrating constant $C_2$ by $\lambda_2=e^{C_2}$. Now, combining the two solutions, we get the complete solution of $G$ in equation \eqref{gf} as follows,
\begin{eqnarray}
\label{gf1}
G(s,t) = f\left[(s-1)e^{-(k_2 + k_3)t}\right]e^{\frac{k_1N+k_4A}{k_2+k_3}s}
\end{eqnarray}
Now, we have to find out the functional form of $f$ in the equation \eqref{gf1} by using the initial condition, $G(s,0)=s^M$ which is given by,
\begin{eqnarray}
\label{ff}
f(s) = (s+1)^Me^{-\frac{k_1N+k_4A}{k_2+k_3}(s+1)}
\end{eqnarray}
Putting this functional form $f(s)$ in equation \eqref{ff} to equation \eqref{gf1} and rearranging the terms, we get the $G(s,t)$,
\begin{eqnarray}
\label{gff}
G(s,t) = \left[(s-1)e^{-(k_2 + k_3)t} + 1\right]^Me^{-\frac{k_1N+k_4A}{k_2+k_3}(1-s)\left[1-e^{-(k_2+k_3)t}\right]}
\end{eqnarray}
Next, we rearrange the terms in the first factor in equation \eqref{gff} and doing the binomial expansion, we have, $\displaystyle\left[(s-1)e^{-(k_2 + k_3)t} + 1\right]^M=\left[se^{-(k_2 + k_3)t} + \left(1-e^{-(k_2 + k_3)t}\right)\right]^M=\sum_i\binom{M}{i}\left[se^{-(k_2 + k_3)t}\right]^i\left[1-e^{-(k_2 + k_3)t}\right]^{M-i}$. From the second factor of the generating function in equation \eqref{gff}, expanding the exponential term having $s$, we have, $\displaystyle e^{\frac{k_1N+k_4A}{k_2+k_3}s\left[1-e^{-(k_2+k_3)t}\right]}=\sum_{j}\frac{1}{j!}\left[\frac{k_1N+k_4A}{k_2+k_3}s\left(1-e^{-(k_2+k_3)t}\right)\right]^j$. Now combining the two expressions in the equation eqref{gff} and rearranging the terms, we have,
\begin{eqnarray}
G(s,t)&=& e^{-\frac{k_1N+k_4A}{k_2+k_3}\left[1-e^{-(k_2+k_3)t}\right]}\sum_i\sum_j\binom{M}{i}\frac{s^{i+j}}{j!}e^{-i(k_2+k_3)t}\left[1-e^{-(k_2+k_3)t}\right]^{M-i}\nonumber\\
&&\times\left[\frac{k_1N+k_4A}{k_2+k_3}(1-e^{-(k_2+k_3)t})\right]^j
\end{eqnarray}
Putting $i+j = x$ in the above expression and comparing with the definition of Generating function $G(s,t) = \sum_{x}s^xP(x,t)$, we finally obtain the probability density function as the following,
\begin{eqnarray}
\label{pd}
P(x,t) = e^{\frac{k_1N+k_4A}{k_2+k_3}\left[1-e^{-(k_2+k_3)t}\right]}\sum_i\binom{M}{i}\left(\frac{k_1N+k_4A}{k_2+k_3}\right)^{x-i}\frac{\left[1-e^{-{(k_2+k_3)t}}\right]^{M+x-2i}}{(x-i)!}e^{-i{(k_2+k_3)t}}
\end{eqnarray}
From this probability distribution \eqref{pd}, it clearly indicates that as $(k_2+k_3)\rangle\rangle (k_1N+k_4A)$ and $(k_2+k_3)\rightarrow large$ limits $e^{-(k_2+k_3)t}\rightarrow 0$, so that $P(x,t)\rightarrow 0$. The situation in this limit indicates that all the available unfolded proteins are converted to folded proteins ($N$) and aggregated unfolded proteins ($A$) as evident from equations \eqref{reaction} and \eqref{pd}. A simple possible condition to control aggregation of unfolded proteins is to find a situation where, $k_2\rangle\rangle k_3;~(k_2+k_3)\rangle\rangle (k_1N+k_4A)$.\\

{\noindent}Now we calculated the observables $\langle x\rangle$ and $\langle x^2\rangle$ which are expectation values of unfolded protein population $x$ using equation \eqref{gff}. Putting $\alpha=\frac{k_1N+k_4A}{k_2+k_3}$ and $\beta=k_2+k_3$, we have,
\begin{eqnarray}
\label{mean}
\langle x\rangle&=&\left.\frac{\partial G(s,t)}{\partial s}\right|_{s=1}\nonumber\\
&=&M\left[(s-1)e^{-\beta t}+1\right]^{M-1}e^{-\beta t}e^{\alpha(1-s)(1-e^{-\beta t})}\nonumber\\
&&+ \left.\left[(s-1)e^{-\beta t}+1\right]^M\left[-\alpha(1-e^{-\beta t})\right]e^{\alpha(1-s)\left[1-e^{-\beta t}\right]}\right|_{s=1}\nonumber\\
&=&\left[M-\alpha(e^{\beta t}-1)\right]e^{-\beta t}
\end{eqnarray}
Further, the second moment of the unfolded protein population $x$ can be obtained as follows,
\begin{eqnarray}
\label{moment}
\langle x^2\rangle &=&\left.\frac{\partial^2 G}{\partial s^2}\right|_{s=1}+\left.\frac{\partial G}{\partial s}\right|_{s=1}\nonumber\\
&=& M(M-1)\left[(s-1)^{e^{-\beta t}} + 1\right]^{M-2}e^{-2\beta t}e^{\alpha (1-s)(1-e^{-\beta t})}\nonumber\\
&&+ M\left[(s-1)^{e^{-\beta t}} + 1\right]^{M-1}e^{-\beta t}(-\alpha\left[1-e^{-\beta t})\right]e^{\alpha (1-s)(1-e^{-\beta t})}\nonumber\\
&&+ M\left[(s-1)^{e^{-\beta t}} + 1\right]^{M-1}e^{-\beta t}\left[-\alpha (1-e^{-\beta t})\right]e^{\alpha (1-s)(1-e^{-\beta t})}\nonumber\\
&&+ \left[(s-1)^{e^{-\beta t}} + 1\right]^M\left[-\alpha(1-e^{-\beta t})\right]^2e^{\alpha (1-s)(1-e^{-\beta t})}\nonumber\\
&&+\langle x\rangle\nonumber\\
&=& M(M-1)e^{-2\beta t} - 2Me^{-\beta t}\alpha (1- e^{-\beta t}) + \left[\alpha(1-e^{-\beta t})\right]^2
\end{eqnarray}
Now, from the equations \eqref{mean} and \eqref{moment}, we arrived at the variance of the population $x$,
\begin{eqnarray}
\label{sigma}
\sigma_x^2&=&\langle x^2\rangle-\langle x\rangle^2\nonumber\\
&=& \left[M-\alpha (e^{\beta t}-1)\right]e^{-\beta t} - Me^{-2\beta t}
\end{eqnarray}

\noindent\textbf{Proposition 1: }\textit{Long time behavior of the population of unfolded protein is given by:
\begin{itemize}
\item Asymptotic limit of the mean $|\langle x\rangle|$ at long time becomes a constant: $\displaystyle \lim_{t\rightarrow\infty}|\langle x\rangle|=\frac{k_1N+k_4A}{k_2+k_3}$.
\item Long time behavior of the $\langle x^2\rangle$ is given by: $\displaystyle \lim_{t\rightarrow\infty}\langle x^2\rangle=\left[\frac{k_1N+k_4A}{k_2+k_3}\right]^2$.
\end{itemize}
}

\noindent\textbf{Proof: }\textit{From the equation \eqref{mean} taking limit $t\rightarrow\infty$, we have,}
\begin{eqnarray}
\lim_{t\rightarrow\infty}|\langle x\rangle|&=&\lim_{t\rightarrow\infty}|\left[M-\alpha(e^{\beta t}-1)\right]e^{-\beta t}|\nonumber\\
&=&\alpha=\frac{k_1N+k_4A}{k_2+k_3}
\end{eqnarray}
\textit{This means that the mean population of the unfolded proteins at the long time limit becomes constant which depends on the rate constants and initial populations of $N$ and $A$. Similarly, by taking the limit $t\rightarrow\infty$ of the equation eqref{moment} we easily get $\displaystyle\lim_{t\rightarrow\infty}\langle x^2\rangle=\left[\frac{k_1N+k_4A}{k_2+k_3}\right]^2$.}\\

\noindent\textbf{Theorem 1: }\textit{In the limit $M,x\rightarrow\infty$, the probability distribution function P(x,t) follows the time dependent Poisson distribution given by,}
\begin{eqnarray}
P(x,t)=F(x,t)Pois(\langle qx\rangle)
\end{eqnarray}
where, $F(x,t)=\displaystyle\Gamma(t)e^{-\frac{M\alpha}{M+\alpha}}e^{\frac{M-\alpha}{M+\alpha}x}\frac{M-\alpha}{M+\alpha}\frac{a}{b}q^{-x}$, $q=\frac{M}{M+\alpha}$.

\noindent\textbf{Proof: }\textit{Let us put, $\displaystyle\Gamma(t)=e^{\frac{k_1N+k_4A}{k_2+k_3}\left[1-e^{-(k_2+k_3)t}\right]}=e^{\alpha\left[1-e^{-\beta t}\right]}$, $p=e^{-(k_2+k_3)t}=e^{-\beta t}$ and rearranging the terms the equation \eqref{pd} can be written as,}
\begin{eqnarray}
\label{pda}
P(x,t)=\Gamma(t)\sum_{i}\binom{M}{i}p^i(1-p)^{M-i}\frac{\alpha^{x-i}(1-p)^{x-i}}{(x-i)!}
\end{eqnarray}
Now in the limit, $M\rightarrow\infty$, we have, $\displaystyle\lim_{M\rightarrow\infty}\binom{M}{i}p^i(1-p)^{M-i}\rightarrow\frac{\lambda^ie^{-\lambda}}{i!},~\lambda=Mp$. Substituting this expression to equation eqref{pfa} and rearranging the term, we have, $P(x,t)=\displaystyle\frac{\Gamma(t)e^{-\lambda}}{x!}\sum_{i}\left[M^i\alpha^{x-i}\right]\frac{x!}{(x-i)!i!}p^i(1-p)^{x-i}$. In the limit $x\rightarrow\infty$, we have, 
\begin{eqnarray}
\label{pxt}
P(x,t)&=&\frac{\Gamma(t)e^{-\lambda}}{x!}\sum_{i}\left[M^i\alpha^{x-i}\right]\frac{\gamma^i e^{-\gamma}}{i!},~~~~~\gamma=xp\nonumber\\
&=&\frac{\Gamma(t)\alpha^{x}e^{-(\lambda+\gamma)}}{x!}\sum_{i}\frac{\left[\frac{M\gamma}{\alpha}\right]^i}{i!}\nonumber\\
&\approx&\frac{\Gamma(t)\alpha^{x}e^{-(M+x)p}}{x!}e^{\frac{Mxp}{\alpha}}\nonumber\\
&=&\Gamma(t)e^{-Mp}\frac{\left[\alpha e^{-p+\frac{Mp}{\alpha}}\right]^x}{x!}
\end{eqnarray} 
From equation \eqref{mean}, we can write first factor in equation eqref{pxt}, $p=\displaystyle\frac{\langle x\rangle+\alpha}{M+\alpha}$. The second factor in the equation \eqref{pxt} can be written as, $e^{-Mp}=e^{-\frac{M\alpha}{M+\alpha}}e^{-\frac{M}{M+\alpha}\langle x\rangle}$. Further, $e^{-p+\frac{M}{\alpha}p}=e^{\frac{M-\alpha}{M+\alpha}x}\left[e^{\frac{1}{\alpha}\frac{M-\alpha}{M+\alpha}\langle x\rangle}\right]^x$. Since, $\alpha <1$ and $M$ is large, we know that $\frac{1}{\alpha}\frac{M-\alpha}{M+\alpha}\langle x\rangle>1$. If we take $z=\frac{1}{\alpha}$, then in this function $\frac{1}{z}e^{z\frac{M-\alpha}{M+\alpha}\langle x\rangle}$, for small values $z$ $1/z$ dominates whereas for large $z$, the exponential function dominates. Now to have approximate solution which have both contributions from both $1/z$ and exponential part is  to consider linear approximation in the functions. For this let us take $\frac{1}{v}=e^{z\frac{M-\alpha}{M+\alpha}\langle x\rangle}<1$, so that, $-z\displaystyle\frac{M-\alpha}{M+\alpha}\langle x\rangle=ln(v)=\sum_{i=0}^{\infty}(-1)^{i}\frac{(q-1)^{i+1}}{i+1}$. Then we consider only the linear terms in the expansion which can be written in the form $av+b\approx -z\displaystyle\frac{M-\alpha}{M+\alpha}\langle x\rangle$, where, $a$ and $b$ are constants, $a,b\in mathcal{R}$. Rewriting this expression again in the form, $[1+\frac{a}{b}v]^{-1}=-\displaystyle\frac{b}{z\frac{M-\alpha}{M+\alpha}\langle x\rangle}$ and collecting only linear terms we have at large $z$ limit, $v=\displaystyle e^{-z\frac{M-\alpha}{M+\alpha}\langle x\rangle}\sim\frac{b}{a}$. Now, from the third factor, for $z\rightarrow\infty$ limit, we have, $\displaystyle\lim_{z\rightarrow\infty}\frac{e^{z\frac{M-\alpha}{M+\alpha}\langle x\rangle}}{z}=\lim_{z\rightarrow\infty}\frac{e^{z\frac{M-\alpha}{M+\alpha}\langle x\rangle}\frac{M-\alpha}{M+\alpha}\langle x\rangle}{1}~(L'Hospital's~rule)\sim\frac{M-\alpha}{M+\alpha}\langle x\rangle\times\frac{a}{b}$. Putting all these results and rearranging the terms, we have,
\begin{eqnarray}
\label{pois}
P(x,t)&\approx& F(x,t)e^{-q\langle x\rangle}\frac{\langle x\rangle^x}{x!}\nonumber\\
&\approx&F(x,t)\left[\frac{e^{-\langle qx\rangle}\langle qx\rangle^x}{x!}\right]\nonumber\\
&\approx&F(x,t)Pois(\langle qx\rangle)
\end{eqnarray}
where, $F(x,t)=\displaystyle\Gamma(t)e^{-\frac{M\alpha}{M+\alpha}}e^{\frac{M-\alpha}{M+\alpha}x}\frac{M-\alpha}{M+\alpha}\frac{a}{b}q^{-x}$, $q=\frac{M}{M+\alpha}$.\\

\noindent\textbf{Theorem 2: }\textit{The asymptotic limiting value of $P(x,t)$ at $M,x,\langle x\rangle\rightarrow\infty$ is a normal distribution given by,}
\begin{eqnarray}
P(x,t)\approx F(x,t)\times N(\langle qx\rangle,\langle qx\rangle)
\end{eqnarray}

\noindent\textbf{Proof: }\textit{Let us take the Poisson distribution in equation \eqref{pois} $Pois(qx)=\displaystyle\frac{e^{-\langle qx\rangle}\langle qx\rangle^x}{x!}$. In $x\rightarrow large$ limit, $x!$ can be approximated by Stirling’s formula, $x!\approx x^x e^{-x}\sqrt{2\pi x}$. Now, one can write, $\displaystyle lnPois(\langle qx\rangle)=xln(\langle qx\rangle)-\langle qx\rangle-xln(x)+x-ln(\sqrt{2\pi x})$. Then $\epsilon-$expansion can be done by expanding $x$ around $\langle qx\rangle$ by, $x=\langle qx\rangle+\epsilon$, where, $\epsilon$ is small arbitrary parameter such that $\displaystyle\frac{\epsilon}{\langle qx\rangle}\langle\langle 1$. Within this approximation and after some algebra, one can reach, $\displaystyle lnPois(\langle qx\rangle)=-\frac{\epsilon^2}{2\langle qx\rangle}-ln\sqrt{2\pi \langle qx\rangle}$. Then we have,}
\begin{eqnarray}
Pois(\langle qx\rangle)&\approx&\frac{1}{\sqrt{2\pi\langle qx\rangle}}e^{-\frac{(x-\langle qx\rangle)^2}{\sqrt{2\langle qx\rangle}}}\nonumber\\
&\approx&N(\langle qx\rangle,\langle qx\rangle)
\end{eqnarray}
Hence, equation eqref{pois} becomes,
\begin{eqnarray}
P(x,t)\approx F(x,t)\times N(\langle qx\rangle,\langle qx\rangle)
\end{eqnarray}

\noindent\textbf{Proposition 2: }\textit{The Fano factor of the distribution of $x$ is given by,}
\begin{eqnarray}
F_f=1-\frac{M}{\langle x\rangle}e^{-\beta t}
\end{eqnarray}
\textit{The processes become:
\begin{itemize}
\item Statistically independent in the limit, $\displaystyle\lim_{t\rightarrow\infty} F_f\rightarrow 1$.
\item Sub-Poissonian process for the case $\displaystyle\frac{M}{\langle x\rangle}e^{-\beta t}>0$ and finite.
\item Noise enhanced process if $\displaystyle\frac{M}{\langle x\rangle}e^{-\beta t}<0$.
\end{itemize}
}

\noindent\textbf{Proof: }\textit{From equations \eqref{mean} and \eqref{sigma}, and using the formula for calculating Fano factor, $F_f=\displaystyle\frac{\sigma_x}{\langle x\rangle}$, which can estimate noise associated with the dynamics \cite{Fano,Chanu}. It is given by,}
\begin{eqnarray}
\label{fan}
F_f=1-\frac{M}{\langle x\rangle}e^{-\beta t}
\end{eqnarray}
\textit{Taking limit $t\rightarrow\infty$ in the equation \eqref{fan}, we have, $F_f=1-\displaystyle\lim_{t\rightarrow\infty}\frac{M}{\langle x\rangle}e^{-\beta t}\rightarrow 1.$ This is the case of Poisson process, where, the processes are statistically independent of each other. }

\noindent\textit{For finite values of $\displaystyle\frac{M}{\langle x\rangle}e^{-\beta t}$, we have $F_f<1$. This is the case of sub-Poissonian process, where, noise does not have significant role in regulating the dynamics.}\\

\noindent\textit{The role of noise in regulating the system's dynamics become visible if the process is noise enhanced or super-Poissonian process. It can be observed if $\frac{M}{\langle x\rangle}e^{-\beta t}<1$. It is possible only when $\beta t<0$ such that $e^{-\beta t}\sim 1-\beta t<0$ satisfying the condition $t>\frac{1}{\beta}$.}\\

{\noindent}Now let us consider the situation of controlling aggregation of unfolded protein, where, $k_2\gg k_3$, the condition at which the rate of converting unfolded to native state of the proteins is much larger than the rate of aggregation of unfolded protein. This condition allows us, $\beta=k_2\displaystyle\left(1+\frac{k_3}{k_2}\right)\sim k_2$ such that, $\displaystyle e^{-\beta t}\sim e^{-k_2t}\sim (1-k_2)t$. This case could be the situation at which unfolded protein is approximately controlled. Then, the Fano factor in equation \eqref{fan} becomes,
\begin{eqnarray}
F_f\sim 1+\frac{M}{\langle x\rangle}(k_2-1)t
\end{eqnarray}
In this situation, the process becomes 1. Poissonian if $F_f=1$; 2. sub-Poissonian if $k_2<1$ and 3. Noise enhanced process if $k_2>1$. Hence, the dynamics as well as distribution of the unfolded protein aggregation is quite different depending on $k_2$ and it can able to drive the dynamics to different dynamical states where processes are drastically different.\\

\noindent\textbf{\large Conclusion}\\
Unfolded protein aggregation is nuisance to the living beings which could be one of the origin of various diseases, namely, \textit{conformational diseases \cite{Gow}, neurodegenerative diseases \cite{Stefani}} etc. This aggregation mechanism is quite complicated process but can be simplified into a simple model we presented which is an extension of the model by Limry and Erying \cite{Lumry}. We considered the model as a Markov process and can be represented by a Masted equation which can be solved using generating function technique by taking certain initial condition. The probability distribution $P(x,t)$ we obtained is quite complicated function which depends on various parameters involved in the model, initial conditions and concentrations of the native protein and aggregated protein. In the long time or aging behavior of the unfolded protein, which can be obtained approximately from the mean value unfolded protein at long time limit, is found to be nearly constant. Since population of such unfolded protein should be minimized to prevent from their aggregation possible condition should be $[k_2+k_3]\gg[k_1N+k_4A]$. This means that total rate of conversion of native state protein and formation of aggregate from unfolded protein should be comparatively larger than the sum of converted unfolded protein populations from native state and aggregated state. Further, conversion from unfolded state to native state should be favored $k_2\gg k_3$.\\

{\noindent}The behavior of the probability distribution $P(x,t)$ at large population of the unfolded limit show modified Poisson distribution with a factor $F(x,t)$ which is mainly dominated by exponential function at large population limit. Further, if we take large limit of mean population of the unfolded protein, the distribution $P(x,t)$ becomes modified normal distribution of same mean and standard deviation. This indicates that the distribution of the interacting unfolded protein is deeply rooted to the universally accepted poisson and normal distributions at various situations. However, the distributions are significantly deviated from the distribution of non-reactively interacting molecules which follow Maxwell-Boltzmann distribution. One reason of the origin of this deviation could be birth and death process involved in the molecular interaction process.\\

{\noindent}The fluctuations inherent to the dynamics of the unfolded protein has significant role in the regulation of the unfolded protein dynamics and has different roles at different situations. This fluctuation parameter can be estimated using standard Fano factor from the mean and variance of the protein dynamics. At large time limit or aging limit of the dynamics indicates that the processes are statistically independent of each other which is the case of completely memory lost dynamics and are not correlated. For finite values of the parameters, $F<1$ indicating possibility of sub-poissinian process where the involved fluctuations try to stabilize the system and hence the system tends to stay at equilibrium state. Noise enhancement process, which generally try to drive the system at non-equilibrium state, could be possible for certain range of time otherwise the dynamics is dominated by sub-poissonian process. Unfolded protein aggregation causes various types of diseases and could be a potential target to prevent from the diseases. The question is how to select aggregation destabilizing molecule for a specific disease. These are open questions which are needed to be investigated systematically.

\vspace{0.5cm}
\noindent{\bf Acknowledgements} \\
{\noindent}We would like to thank Jawaharlal Nehru University for providing us facilities to carry out this work.
\\

\noindent {\bf Author Contributions:}\\
{\noindent}RKBS conceptualized and supervised the work. UU, CB, SD, JPK and RKBS did the analytical work. UU, CB, SD, JPK and RKBS wrote the manuscript. All the authors read and approved the manuscript.\\

\noindent {\bf Additional Information} \\
\textbf{Competing financial interests:} The authors declare no competing interests.

\end{document}